\documentclass{mem}
\usepackage{natbib}
\usepackage{txfonts}
\usepackage{balance}
\usepackage{graphicx}
\usepackage{flushend}
\usepackage[breaklinks,pdftex]{hyperref}
\idline{75}{282}
\begin{document}
\def\teff{$T\rm_{eff }$}
\def\kms{$\mathrm {km s}^{-1}$}

\title{
Recent Highlights from the VERITAS AGN Discovery Program 
}

\author{
W. \,Benbow\inst{1} 
\and J. \,Christiansen\inst{2}
\and J. Francescutti\inst{2}
\and G. Kunkler\inst{2} 
\and W. Root\inst{2}
\and P. Zyla\inst{2}
\and the VERITAS Collaboration\inst{3}
          }

\institute{
Center for Astrophysics $|$ Harvard \& Smithsonian, Cambridge, MA 02138, USA
\and
 Physics Department, California Polytechnic State University, San Luis Obispo, CA 94307, USA
\and
https://veritas.sao.arizona.edu/\\
\email{wbenbow@cfa.harvard.edu}
}

\authorrunning{Benbow}

\titlerunning{VERITAS AGN Discovery Program}

\date{Received: Day Month Year; Accepted: Day Month Year}

\abstract{
VERITAS began full-scale operations in 2007 and it remains one of the world’s most sensitive very-high-energy (VHE; E $>$ 100 GeV) gamma-ray observatories. More than 8,300 hours ($\sim$50\%) of its good-weather data were targeted on active galactic nuclei (AGN). Many of these observations were taken as part of an ongoing comprehensive program to discover new VHE AGN. Upon discovery, the VERITAS collaboration leverages VHE spectral and variability measurements, and accompanying broadband observations to probe the underlying jet-powered processes in AGN. Recent scientific highlights from the VERITAS AGN discovery program, including the VHE discoveries of B2\,0912+29, 1ES\,1028+511, 1ES\,1118+424 and RBS\,1366, are presented.
\keywords{Gamma-ray Astronomy, Active Galactic Nuclei, Blazars, Gamma-ray Sources}
}
\maketitle{}

\section{Introduction} 
AGN rank among the most powerful particle accelerators in the universe 
and are a major component of the observing programs of VHE observatories
such as VERITAS \citep{VERITAS_paper}. These objects are found at the center of some
galaxies and are related to their central supermassive black holes (SMBHs).
They are the most numerous class of identified VHE $\gamma$-ray sources
and comprise $\sim$30\% of the VHE catalog \citep{TEVCAT}.

A small fraction of AGN possess strong collimated outflows (jets) powered by accretion onto the SMBH.
VHE $\gamma$-ray emission from AGN is believed to be produced in these relativistic jets, in a compact region
near the SMBH. However, there is also evidence that VHE emission can be 
produced on larger and/or more distant scales (see, e.g., \cite{VERITAS_1441}). 

The spectral energy distributions (SEDs) observed from AGN are highly non-thermal,
span from radio waves through $\gamma$-rays, and are 
characterized by a double-peaked structure.  Rapid 
multi-wavelength (MWL) variations, on time scales as short as minutes, 
are commonly observed from these variable sources.  Accordingly,
contemporaneous MWL observations are a key component in probing the 
physical mechanisms powering these objects. In general, the lower-energy SED peak 
in AGN is attributed to synchrotron radiation from relativistic electrons in the jet. 
Although hadronic models are still feasible (e.g.,~\cite{Cerruti-UHBL}), 
the higher-energy SED peak is commonly believed to be due to inverse-Compton (IC) scattering, 
with the origin of the seed photons being either from the synchrotron process 
(synchrotron self-Compton (SSC)) or from 
external sources (external Compton (EC)) \citep{ghisellini1998theoretical}.

Blazars are the dominant class ($\sim$93\%) of VHE AGN, and are those AGN
with jets pointed along the line of sight towards Earth.
Of the 87 VHE blazars, at least 71 ($\sim$80\%) are BL Lac objects and
10 are Flat Spectrum Radio Quasars (FSRQs); 6 have uncertain sub-classification.
The jets for the remaining 7 VHE AGN are not strongly misaligned.
At least four of these AGN are nearby ($z < 0.022$) FR-I radio galaxies. 
The VHE AGN catalog is peaked at nearby redshifts (e.g., $\sim$55\% have $z < 0.2$)
and currently covers a redshift range from $z = 0.0018$ to $z = 0.997$.  However, it
is important to note that at least $\sim$20\% of VHE AGN have unknown redshift 
and many catalog redshifts are uncertain due to the defining absence of optical features 
in the spectra of BL Lac objects. The observed 
redshift distribution is largely the result of 
the strong attenuation of VHE photons
by the extragalactic background light (EBL) \citep{VERITAS_EBL}.

\vspace{-0.2cm}
\section{VERITAS AGN Program}

VERITAS performs studies of astrophysical sources between $\sim$100 GeV and $\sim$30 TeV.  
Approximately 19,400 h of observations have been acquired since VERITAS
began full-scale operations in 2007.  Currently, the project averages $\sim$910 h / yr and
$\sim$140 h / yr of good-weather dark-time and bright-moon time, respectively.  The latter has similar sensitivity to dark-time observations above $\sim$250 GeV.
VERITAS detects a 1\% Crab Nebula flux source in $<$25 h of observations and the
typical systematic errors for its measurements are 20\% on the flux and 0.1 on the
photon index.

The VERITAS AGN (radio galaxy and blazar) Program comprises $\sim$50\% of its observations.
Between 2007-24, $\sim$7,000 h of good-weather dark time ($\sim$410 h / yr)  were dedicated
to AGN.  Historically this was split 90\% blazars and 10\% radio galaxies, but the radio galaxy share
has recently grown to 25\% of the AGN dark time observations.  The bright-moon observations
began in 2012, and since then $\sim$1,350 h ($\sim$110 h / yr) were dedicated to blazars.
Generally, VERITAS blazar observations are focused on BL\,Lac objects. The AGN program has 5 themes:
\vspace{-0.1cm}
\begin{description}
 \item[$\bullet$] {\bf Discovery of new blazars at VHE:} This is $\sim$35\% of the program and is described in detail in Section~\ref{Discovery}.  

 \item[$\bullet$] {\bf Target-of-Opportunity (ToO) observations:} Significant, high-priority time is used to respond to 
both self-triggers and flare alerts from MWL and multi-messenger instruments.
Any target ``of interest" from IceCube or exceptional VHE
flare receives intense coverage.
 
 \item[$\bullet$] {\bf Monitoring of bright VHE AGN:}. Regular, long-term monitoring of particularly interesting VHE AGN 
is used to self-trigger intense MWL campaigns, expand our long-term light curves and 
build high-statistics spectra for a variety of AGN sub-types in a range of states.
 
 \item[$\bullet$] {\bf Simultaneous MWL observations:} Studies leveraging contemporaneous MWL observations 
are essential for constraining the sites of emission, particle species 
and radiation processes in AGN jets.  VERITAS exploits synergies
with {\emph Fermi}-LAT, Swift-XRT, IXPE, NuSTAR and EHT.
 
 \item[$\bullet$] {\bf Observations of radio galaxies:} The radio galaxy program 
 is $\sim$65\% focused on the discovery of new VHE emitters, and 
$\sim$35\% focused on better characterizing the SEDs of known VHE emitters.
\end{description}

\vspace{-0.4cm}
\section{VERITAS AGN Discovery Program \label{Discovery}}

The discovery of new AGN is a key part of the VERITAS science program.
Previous studies yield considerable understanding of VHE AGN
and their related science. The VERITAS program seeks to further
develop these conclusions based on a larger population, 
from different sub-classes, and with contemporaneous SED modeling in all cases.

There are currently 94 active galactic nuclei (AGN) detected at VHE, 
and these sources are dominated ($\sim$80\%) by BL Lac objects. Population studies 
\citep{ackermann2011second,nolan2012fermi} indicate a correlation between synchrotron and 
IC peak frequencies, and perhaps unsurprisingly $\sim$80\% of the BL Lacs detected at 
VHE are high-frequency-peaked (HBLs; i.e. those with synchrotron peak 
frequency, $\nu_{peak}$, above $\sim$$10^{15}$ Hz).  Excepting radio galaxies, which are
generally weakly variable, almost all of the non-HBLs detected in the VHE
band were detected during flares.  Accordingly the VERITAS
AGN discovery program has 2 approaches: pre-planned observations 
of HBLs and radio galaxies
and Target-of-Opportunity (ToO) observations of AGN from all classes.  The
VERITAS pre-planned program 
has evolved over time. Briefly, it is a comprehesive
survey of the X-ray brightest 2WHSP HBLs (TeV FOM $>$ 1.0) 
and the hardest 2FHL objects ($\Gamma_{2FHL} < 2.8$).  Other targets were observed, particularly $Fermi$-LAT motivated radio galaxies
and IBLs. For the ToO program, VERITAS exploits
a variety of triggers including those from $\gamma$-ray,
multi-messenger, and optical facilities.  Excluding 17 pre-2023 discoveries,
the AGN discovery program emposasses $\sim$1,600 h of good-weather observations
on $\sim$185 targets.  As every target in the comprehensive survey
has some VERITAS exposure, the program now emphasizes follow-up on weak ($>$3$\sigma$) excesses seen in archival exposures.

\begin{table*}[t]
\caption{Key MWL traits of four recent VERITAS AGN discoveries.  Shown are the redshift, TeV FoM \citep{3HSP}, logarithm of the synchrotron peak frequency, $\Gamma$ in the 4FGL (8 yr, 50 MeV - 1 TeV), 3FHL (7 yr, $>10$ GeV) and 2FHL (6.7 yr, $>50$ GeV) catalogs, and the fluxes from 2FHL ($>50$ GeV) and from 3FHL (50 - 150 GeV).}
\vspace{-0.8cm}
\label{SourceSummary}
\begin{center}
\resizebox{\hsize}{!}{
\begin{tabular}{c | c c c | c c c c c}
\hline\hline
AGN & z & FoM & $\nu_{peak}$ &  $\Gamma_{4FGL}$  & $\Gamma_{3FHL}$ & $\Gamma_{2FHL}$ & $\Phi (> 50 ~GeV)$ & $\Phi (50 - 150 ~GeV)$ \\
\hline
B2\,0912+29 & $>$0.19 & 3.16 & 15.4 & $ 1.88 \pm 0.02 $ & $ 2.39 \pm 0.18 $ & $ 4.5 \pm 2.0 $ & $2.5\%$  Crab &  $3.2\%$ Crab \\
1ES\,1028+511 & 0.361 & 3.98 & 16.9 & $ 1.76 \pm 0.04 $ & $ 1.94 \pm 0.14 $ & $ 2.6 \pm 0.6 $ & $2.5\%$  Crab & $1.0\%$  Crab \\
1ES\,1118+424 &  $>$0.28 & 2.00 & 16.2 & $ 1.62 \pm 0.06 $ & $ 2.07 \pm 0.11 $ & $ 2.5 \pm 0.5 $ & $3.7\%$  Crab & $3.4\%$  Crab \\
RBS\,1366 &  0.237 & 3.16 & 17.6 & $ 1.50 \pm 0.07 $ & $ 1.87 \pm 0.23 $ & $ 2.4 \pm 0.7 $ & $1.3\%$  Crab & $1.3\%$  Crab \\
\hline\hline
\end{tabular}
}
\end{center}
\vspace{-0.5cm}
\end{table*}

\vspace{-0.4cm}
\section{Recent Highlights}
Four blazars were discovered at VHE by VERITAS since 2022.  
Each of these HBLs are relatively distant (z $\gtrsim$ 0.2)
and well-studied at MWL.  All are
members of the 3HSP blazar catalog \citep{3HSP} and the recent 
major $Fermi$-LAT catalogs: 4FGL (50 MeV - 1 TeV) \citep{4FGL}, 
2FHL ($\geq$50 GeV) \citep{2FHL} and 3FHL  ($\geq$10 GeV) \citep{3FHL}.
Table~\ref{SourceSummary} summarizes key MWL characteristics of these HBLs. Of particular note is the 3HSP TeV Figure of Merit (FoM) parameter which indicates an object's synchrotron-peak flux; 
it is the ratio of this flux to that of the weakest source detected at VHE. Assuming the SSC model applies, and FoM $>$ 1 implies a blazar should be detectable at VHE. 

\begin{figure*}[t]
\begin{center}
\resizebox{0.8\hsize}{!}{\includegraphics[clip=true]{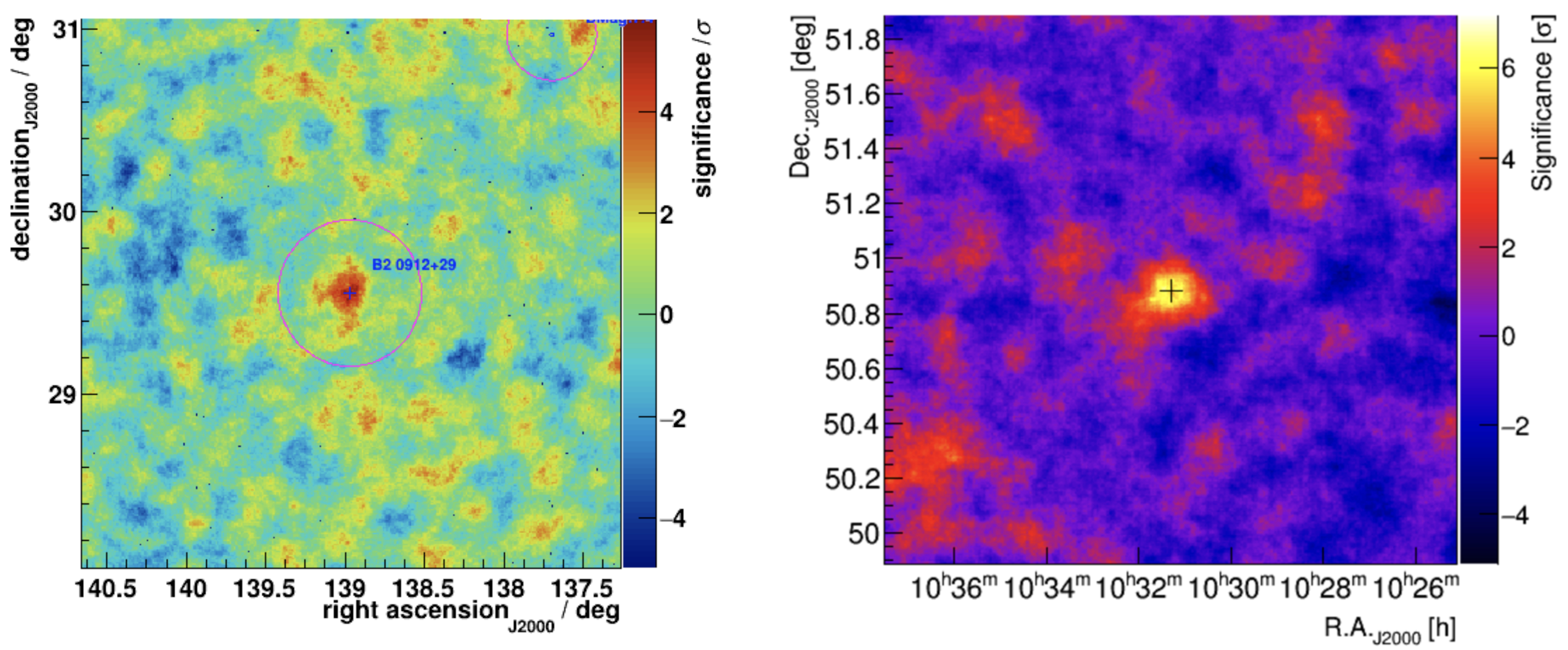}}
\resizebox{0.8\hsize}{!}{\includegraphics[clip=true]{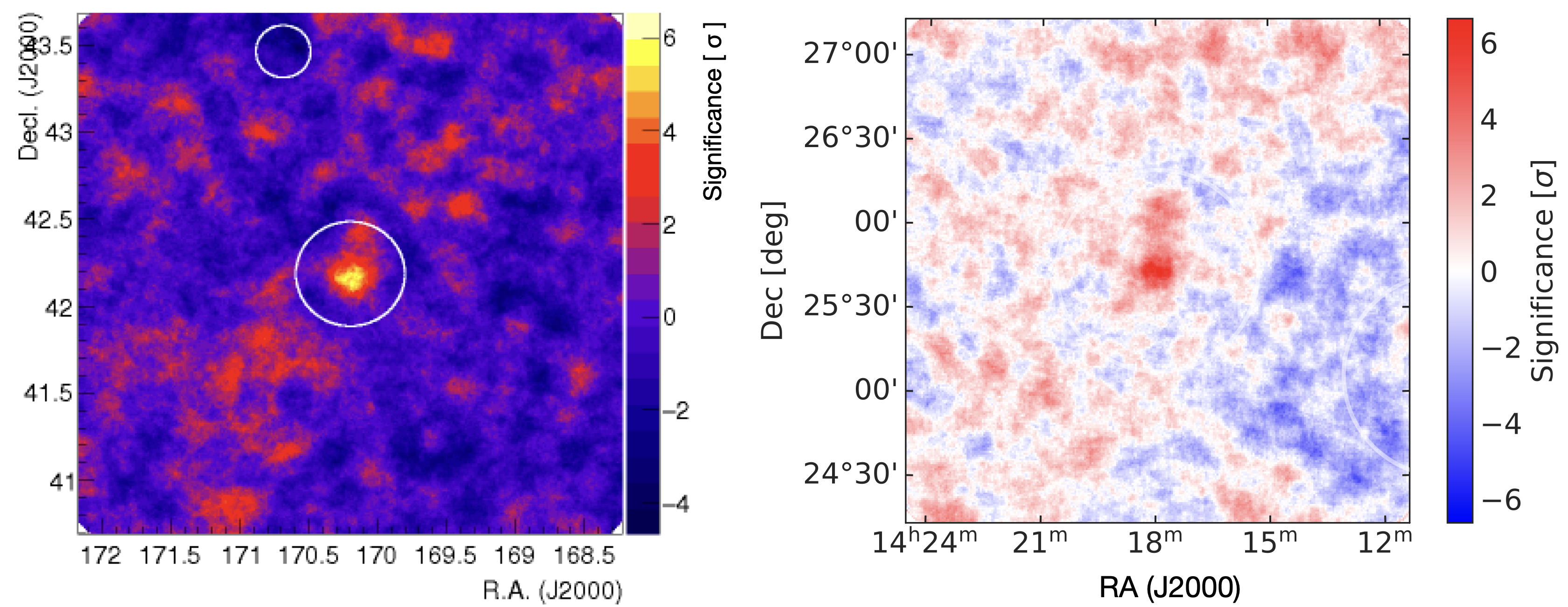}}
\caption{\footnotesize Preliminary significance maps from VERITAS observations of (Upper Left) B2\,0912+29, (Upper Right) 1ES\,1028+511, (Lower Left) 1ES\,1118+424 and (Lower Right) RBS\,1366.}
\label{results_panel1}
\vspace{-0.8cm}
\end{center}
\end{figure*}

{\bf B2\,0912+29} is a borderline IBL/HBL that is moderately variable in Swift-XRT and $Fermi$-LAT data. 
It is one of the brightest objects in the ROXA \citep{ROXA} 
hard-blazar survey and 
was one of the most promising VHE targets initially reported by $Fermi$-LAT (1FGL) \citep{1FGL}. 
Approximately 62 hours of quality-selected exposure were acquired on B2\,0912+29 from 2011 to 2024.
A preliminary analysis of these data yields a detection (322 $\gamma$-rays,
$\sim$5.8 standard deviations, $\sigma$). A sky map of the significance observed 
near B2\,0912+29 is shown in Figure~\ref{results_panel1}.  
The observed VHE photon spectrum is soft ($\Gamma$ = 4.3 $\pm$ 0.6) and the VHE flux is constant, F($>$ 200 GeV) = $(1.3 \pm 0.3_{\mathrm stat}) \times 10^{-12}$ cm$^{-2}$ s$^{-1}$.
This is approximately 0.5\% of the Crab Nebula flux (Crab) 
above the same threshold. 
Thirty-two archival Swift UVOT / XRT exposures exist, including 15 from a VERITAS-initiated campaign in 2013
where the X-ray flux was lower than at other times.

{\bf 1ES\,1028+511} is an extreme HBL (EHBL) with a steady \textit{Fermi}-LAT flux.
It has 33 archival Swift UVOT / XRT exposures, and factor of 3 XRT count-rate variations are seen. It was initially suggested as a candidate VHE emitter based on its 
optical / X-ray SED \citep{luigi_AGN}.  1ES\,1028+511 is one of the brightest objects in both the Sedentary \citep{Sedentary} (extreme HBL) 
and ROXA \citep{ROXA} surveys, and its TeV FoM ($\sim$4) is the 35$^{th}$ brightest in 3HSP. 
A significant cluster of VHE photons was found at the source location in 
\textit{Fermi}-LAT data \citep{Armstrong2015}, 
and the object was observed at $\sim$3.6 sigma between 150 and 500 GeV in the 
3FHL catalog \citep{3FHL}.  
1ES\,1028+511 was observed for $\sim$49 h of quality-selected
live time between 2007 and 2024. A preliminary analysis of these data yields a
VHE detection (6.1$\sigma$, 320 $\gamma$-rays). 
The significance map for the region 
near 1ES\,1028+511 is shown in Figure~\ref{results_panel1}. 
The observed photon spectrum is soft 
($\Gamma = 3.6 \pm 0.5$). The corresponding
flux is F($>$ 200 GeV) = $(2.4 \pm 0.5_{\mathrm stat}) \times 10^{-12}$ cm$^{-2}$ s$^{-1}$ or $\sim$1.0\% Crab, and 
is similar to the 3FHL value. 

{\bf 1ES\,1118+424} was initially suggested to be a VHE emitter because it was a bright Einstein HBL \citep{stecker_AGN}. It is a variable
\textit{Fermi}-LAT source, and its flux between 150 and 500 GeV
is significant (6.1$\sigma$, 3\% Crab) in the 3FHL catalog. 
An analysis also shows a significant cluster of VHE photons \citep{Armstrong2015}, 
corresponding to $\sim$3-4\% Crab, consistent with the 2FHL / 3FHL values in Table~\ref{SourceSummary}. 
Approximately 40 h of quality-selected
observations of 1ES\,1118+424 were acquired between 2007 and 2024.
A preliminary analysis of these data yields a soft-spectrum ($\Gamma \sim 3.9$) VHE detection (5.8$\sigma$, 247 
$\gamma$-rays). A significance map of the region 
near 1ES\,1118+424 is shown in Figure~\ref{results_panel1}. 
The observed flux F($>$ 200 GeV) = $(1.9 \pm 0.4_{\mathrm stat}) \times 10^{-12}$ cm$^{-2}$ s$^{-1}$ is approximately 0.8\% Crab, and lower than the fluxes
from the \textit{Fermi}-LAT analyses. At MWL, 21 Swift UVOT / XRT exposures exist, many of which are simultaneous to VERITAS data. These show factor of 
$\sim$3 variations in the XRT count rates and
indicate a lower flux during the VERITAS observations.

{\bf RBS\,1366} is another EHBL with a 
steady \textit{Fermi}-LAT flux.  Similar to
1ES\,1028+511 it was initially suggested to be a VHE emitter based on its 
optical / X-ray SED \citep{luigi_AGN}. It also one of 
the brighter objects in the Sedentary, ROXA, and 3HSP catalogs.
Approximately 80 hours of quality-selected exposure were acquired on 
RBS\,1366 between 2008 and 2024. A preliminary analysis of 
these data yields a strong VHE detection (440 $\gamma$-rays,
$\sim$6.9$\sigma$) corresponding to
a steady flux of F($>$ 200 GeV) $\sim$ 0.7\% of the Crab.
A sky map of the significance observed 
near RBS\,1366 is shown in Figure~\ref{results_panel1}.
The observed VHE photon spectrum from RBS\,1366 is unusually hard
($\Gamma$ = 2.7 $\pm$ 0.5). 
Combining the \textit{Fermi}-LAT spectrum, with the EBL-corrected
VHE data, yields an one of the hardest spectra ($\Gamma \sim 1.6$) observed from
an EHBL, making it an interesting source for modeling efforts.

{\bf IC\,310} is a head-tail radio galaxy at redshift z = 0.0189, and a known
VHE emitter. It is generally characterized by a low VHE flux, but
has previously flared ($\sim$50\% Crab) with 10-minute variability time scales.  The object
is detected by VERITAS during extensive observations from
2009-17 ($\sim$7$\sigma$ in 80 h; $\sim$1\% Crab).  In March 2024, the LHAASO collaboration
reported several bright flares of IC\,310 (e.g., $\sim$50\% Crab; ATel\#16513).
VERITAS performed follow-up observations of these flares, at large zenith angle,
resulting in a detection ($\sim$7$\sigma$ in 2 h; $\sim$15\% Crab). The success
of the follow-up campaign
demonstrates that the LHAASO all-sky monitor 
can successfully trigger VERITAS on VHE flares of otherwise dim AGN.
This adds a new instrument to the suite used for discovery ToO triggers.

\vspace{-0.4cm}
\section{Final Remarks}
VERITAS has completed 17 seasons of operations and continues to run very well.
The collaboration plans to operate the observatory through at least
2028. The VERITAS catalog consists of 69 sources, belonging
to 8 classes. Of these 45 are AGN, and $\sim$45\% of 
these were VHE discoveries. A 
comprehensive survey of the hardest 2FHL and brightest 2WHSP objects is complete, and the collaboration is now
following up on weak excesses (3-5$\sigma$) from this effort. The four discoveries presented here
are the result of this ongoing follow-up effort.  The VERITAS discovery ToO program will also continue.  Upper limits from the 
$\sim$180 non-detected objects 
are in preparation.

\begin{acknowledgements}
This research is supported by grants from the U.S. Department of Energy Office of Science, the U.S. National Science Foundation and the Smithsonian Institution, by NSERC in Canada, and by the Helmholtz Association in Germany. We acknowledge the excellent work of the technical support staff at the Fred Lawrence Whipple Observatory and at the collaborating institutions in the construction and operation of the instrument.
\end{acknowledgements}

\bibliographystyle{aa}
\bibliography{bibliography}

\end{document}